\documentclass{article}

\usepackage{amsmath,eufrak,eucal}

\newcommand\comment[1]{} 
\newcommand{\eprint}{\textsf} 

\newcommand\journalfont{\rm}  
\newcommand\jou[1]{{\journalfont #1\ }}
\newcommand\joudef[2]{\newcommand #1{\jou{\ignorespaces #2}}}

\joudef{\aaa}    { Astron.\ Astrophys.}
\joudef{\aip}    { Adv.\ Phys.}
\joudef{\adm}    { Adv.\ Math.}
\joudef{\am}     { Ann.\ Math.}
\joudef{\apny}   { Ann.\ Phys.\ (N.Y.)}
\joudef{\apj}    { Astrophys.\ J.}
\joudef{\apjs}   { Astrophys.\ J.\ Suppl.}
\joudef{\cjp}    { Can.\ J.\ Phys.}
\joudef{\cmp}    { Commun.\ Math.\ Phys.}
\joudef{\cqg}    { Class.\ Quantum Grav.}
\joudef{\faa}    { Funct.\ Anal.\ Appl.}
\joudef{\grg}    { Gen.\ Rel.\ Grav.}
\joudef{\ijmpd}  { Int.\ J.\ Mod.\ Phys.\ D}
\joudef{\ijtp}   { Int.\ J.\ Theor.\ Phys.}
\joudef{\invm}   { Invent.\ Math.}
\joudef{\jm}     { J.\ Math.}
\joudef{\jmp}    { J.\ Math.\ Phys.}
\joudef{\jpa}    { J.\ Phys.\ A}
\joudef{\mnras}  { Mon.\ Not.\ R.\ Ast.\ Soc.}
\joudef{\mpla}   { Mod.\ Phys.\ Lett.\ A} 
\joudef{\nature} { Nature}
\joudef{\nc}     { Nuovo Cim.}
\joudef{\npb}    { Nuc.\ Phys.\ B}
\joudef{\ph}     { Physica}
\joudef{\pla}    { Phys.\ Lett. A}
\joudef{\plb}    { Phys.\ Lett. B}
\joudef{\pr}     { Phys.\ Rev.}
\joudef{\prd}    { Phys.\ Rev.\ D}
\joudef{\prep}   { Phys.\ Rep.}
\joudef{\prl}    { Phys.\ Rev.\ Lett.}
\joudef{\prsla}  { Proc.\ Roy.\ Soc.\ Lond.\ A}
\joudef{\ptp}    { Prog.\ Theor.\ Phys.}
\joudef{\ptps}   { Prog.\ Theor.\ Phys.\ Suppl.}
\joudef\rmp      { Rev.\ Mod.\ Phys.}
\joudef\spj      { Sov.\ Phys.\ JETP}

\catcode`@=11

\newcommand\eqalign[1]{\null\,\vcenter{\openup\jot\m@th
  \ialign{\strut\hfil$\displaystyle{##}$&$\displaystyle{{}##}$\hfil
      \crcr#1\crcr}}\,}
\newcommand\meqalign[1]{\null\,\vcenter{\openup\jot\m@th
  \ialign{\strut\hfil$\displaystyle{##}$&&$\displaystyle{{}##}$\hfil
      \crcr#1\crcr}}\,}
\def\ps@reportnumber{%
    \let\@oddfoot\@empty\let\@evenfoot\@empty
    \def\@oddhead{\hfil\rightmark}}
	
\catcode`@=12   

\newdimen\arrayruleHwidth
\makeatletter
\newcommand\Hline{\noalign{\ifnum0=`}\fi\hrule \@height \arrayruleHwidth
  \futurelet \@tempa\@xhline}
\makeatother

\newcommand\thickbaselines{\baselineskip=20pt\lineskip=3pt\lineskiplimit=3pt}

\catcode`@=11
\renewcommand\cases[1]{\left\{\,\vcenter{\thickbaselines\m@th
             \ialign{$##\hfil$&\quad##\hfil\crcr#1\crcr}}\right.}
\renewcommand\matrix[1]{\null\,\vcenter{\thickbaselines\m@th
    \ialign{\hfil$##$\hfil&&\quad\hfil$##$\hfil\crcr
      \mathstrut\crcr\noalign{\kern-\baselineskip}
      #1\crcr\mathstrut\crcr\noalign{\kern-\baselineskip}}}\,} 
\catcode`@=12   

\newcommand\be{\begin{equation}} \newcommand\ee{\end{equation}} 
\newcommand\bd{\begin{displaymath}}\newcommand\ed{\end{displaymath}}

\newcommand\Tr{\mathop{\rm Tr}\nolimits}

 \newcommand\Rscr{{\cal R}}
 
 \newcommand\Vscr{{\cal V}}

\newcommand\undersim[1]{\mathop{\vtop{\ialign{##\crcr
     $\hfil\displaystyle{#1}\hfil$\crcr\noalign
     {\kern1pt\nointerlineskip}\hbox{$\hfil\sim\hfil$}\crcr
     \noalign{\kern1pt}}}}}

\newcommand\case[2]{\textstyle{\frac{#1}{#2}}}

\newcommand{\acronym}[3]{\newcommand{#1}{#3 (#2)\relax\renewcommand{#1}{#2}}}

\acronym{\cybe}{CYBE}{classical Yang-Baxter equation}

\newcommand{\del}{\partial}

\newcommand{\bo}{\boldsymbol}
\newcommand\bfA{{\bo A}} \newcommand\bfB{{\bo B}}
 
  \newcommand\bfE{{\bo E}}
 \newcommand\bfF{{\bo F}} 
 \newcommand\bfL{{\bo L}} 
  
\newcommand\bfr{{\bo r}} 
\newcommand\bfs{{\bo s}} 
\newcommand\bfT{{\bo T}} \newcommand\bfX{{\bo X}}
\newcommand\bfu{{\bo u}} \newcommand\bfU{{\bo U}}
\newcommand\bfw{{\bo w}} 
\newcommand\bfY{{\bo Y}}

\newcommand\bfGamma{\boldsymbol{\Gamma}}

\newcommand\bfRscr{\boldsymbol{\Rscr}}

\newcommand\id{{\bo 1}}

\newcommand{\rbigl}[1]{\raisebox{.35ex}{$\bigl#1$}} 
\newcommand{\rbigr}[1]{\raisebox{.35ex}{$\bigr#1$}} 

\newcommand{\bs}[2]{\overset{{\scriptscriptstyle(#2)}}{{\boldsymbol{#1}}}}
\newcommand{\bsL}[2]{\overset{\!\!\!\!\!\scriptscriptstyle(#1)}{\bfL^{#2}}}

\begin{document}
\bibliographystyle{prsty}

\title{The classical $r$-matrix in a geometric framework
       \footnote{\textsf{solv-int/9801017}}}
\author{Kjell Rosquist \\
        {\small Department of Physics, Stockholm University}  \\[-10pt]
        {\small Box 6730, 113 85 Stockholm, Sweden} \\
        {\small E-mail: \textsf{kr@physto.se}}}
\date{}
\maketitle
\thispagestyle{reportnumber}\markright{\hspace{15.5cm}USITP 98-01}

\begin{abstract}
We use a Riemannian (or pseudo-Riemannian) geometric framework to formulate 
the theory of the classical $r$-matrix for integrable systems.  In this 
picture the $r$-matrix is related to a fourth rank tensor, named the {\em 
$r$-tensor}, on the configuration space.  The $r$-matrix itself carries one 
connection type index and three tensorial indices.  Being defined on the 
configuration space it has no momentum dependence but is dynamical in the 
sense of depending on the configuration variables.  The tensorial nature of 
the $r$-matrix is used to derive its transformation properties.  The resulting 
transformation formula turns out to be valid for a general $r$-matrix 
structure independently of the geometric framework.  Moreover, the entire 
structure of the $r$-matrix equation follows directly from a simple covariant 
expression involving the Lax matrix and its covariant derivative.  Therefore 
it is argued that the geometric formulation proposed here helps to improve the 
understanding of general $r$-matrix structures.  It is also shown how the 
Jacobi identity gives rise to a generalized dynamical classical Yang-Baxter 
equation involving the Riemannian curvature.
\end{abstract}

\section{Introduction}

The classical $r$-matrix is a fundamental object in the study of classically 
integrable systems \cite{semenov-tian-shansky:rmatrix,bv:hamlax}.  It appears 
at an even more fundamental level than the equally important Lax pair 
structure.  However, the nature of the classical $r$-matrix has remained 
somewhat obscure for almost two decades.  For example, some $r$-matrices are 
purely numerical while others depend on the configuration variables (see 
\cite{sklyanin:dynamicalr} and references therein) and there are also examples 
where the $r$-matrix depends on the momentum variables 
\cite{braden:conjecturerm}.  The purpose of this paper is to contribute 
towards a better understanding of the classical $r$-matrix.  In particular we 
demonstrate how a general $r$-matrix can be transformed by a coordinate 
transformation.  The transformation formula is explained in terms of a 
geometric interpretation of the $r$-matrix.  In this picture, the $r$-matrix 
appears as a four-index object carrying three tensorial indices and one 
connection type index.  The formulation given here relies on the geometric 
formulation proposed earlier \cite{rosquist:lax,rg:intspacetimes} for the Lax 
pair equation $\dot\bfL = [\bfL, \bfA]$ (we use boldface symbols for matrices 
throughout the paper).  In that formulation, the Lax matrix itself ($\bfL$) is 
related to a tensor carrying three indices, called the (first) Lax tensor 
while the second Lax matrix ($\bfA$) is related to a connection type object.  
In \cite{rosquist:lax} a fully covariant formulation was achieved in which the 
second Lax matrix was replaced by an object (called the second Lax tensor) of 
the same tensorial type as the first Lax tensor.  Similarly, in the geometric 
version of the $r$-matrix structure to be outlined in this letter the 
$r$-matrix is replaced by a fourth rank tensorial object, named the 
$r$-tensor.  We now describe this procedure starting with the already known 
tensorial Lax pair formulation as given in \cite{rosquist:lax}.  A more 
detailed account of this work will be published elsewhere 
\cite{rosquist:rmatrix}.

\section{The covariant Lax pair equation}
The approach taken here is purely differential geometric without any reference 
to an underlying group theoretic structure.  Also, for simplicity, we use a 
formulation without spectral parameters.  However, we expect that the 
fundamental conclusions will remain valid also in the presence of spectral 
parameters.  It is assumed that the dynamical system under study is described 
in terms of the phase space variables $(q^\alpha,p_\beta)$ labelled by greek 
indices, $\alpha,\beta,\ldots = 1,2,\ldots,d$.  The equations of motion are 
taken to be the geodesics with respect to a Riemannian or pseudo-Riemannian 
metric
\begin{equation}\label{eq:metric}
   ds^2 = g_{\alpha\beta}(q) \, dq^\alpha dq^\beta \ .
\end{equation}
The system can then be represented by the Hamiltonian
\begin{equation}\label{eq:ham}
   H = \case12 g^{\alpha\beta} p_\alpha p_\beta \ ,
\end{equation}
where $g^{\alpha\beta}$ is the matrix inverse of $g_{\alpha\beta}$.  The 
application of this kind of geometric formalism is however not restricted to 
purely kinetic Hamiltonians.  In fact it is well-known that any Hamiltonian 
system of the form $H = \case12 g^{\alpha\beta} p_\alpha p_\beta + U(q)$ can 
be put in the form \eqref{eq:ham} by a change of time variable (see {\em e.g.} 
\cite{arnold:mechanics,lanczos:mechanics} and {\em cf.} also 
\cite{rg:intspacetimes} for an alternative geometrization scheme using 
canonical transformations).  To arrive at the geometric formulation of the Lax 
pair equation
\begin{equation}\label{eq:lax}
   \dot \bfL := \{\bfL, H\} = [\bfL, \bfA] \ ,
\end{equation}
we write the Lax pair matrices, $\bfL = (L^\alpha{}_\beta)$, $\bfA = 
(A^\alpha{}_\beta)$ using mixed indices with contravariant (up) indices 
labelling rows and covariant (down) indices labelling columns.  This notation 
is necessary in order to make matrix multiplication a covariant operation.  
However, it is in fact very convenient in the noncovariant picture as well.  
For the covariant formulation we also need the matrices defined by $\bfL^\mu 
:= \partial\bfL/ \partial p_\mu = (L^\alpha{}_\beta{}^\mu)$ and $\bfA^\mu := 
\partial\bfA/ \partial p_\mu = (A^\alpha{}_\beta{}^\mu)$.

It is assumed in this paper that the Lax matrices in the geometric picture are 
linear and homogeneous in the momenta.  Although this does not give the most 
general geometrization framework it is the simplest and perhaps most elegant 
subcase.  Also, as shown in \cite{rg:intspacetimes}, the 3-particle 
nonperiodic Toda lattice can be given such a linear geometric Lax formulation.  
Furthermore, it turns out that an important application given in the present 
letter, namely the transformation properties of the $r$-matrix, will be seen 
to be valid independently of this assumption.  The matrices $\bfL^\mu$ and 
$\bfA^\mu$ are then functions on the configuration space.  The object with 
components $L^\alpha{}_\beta{}^\mu$ will be interpreted as a third rank 
tensor, the {\em first Lax tensor}.  The {\em second Lax tensor\/} has 
components $B^\alpha{}_\beta{}^\mu$ and is defined by $\bfB^\mu := 
\partial\bfB/ p_\mu = (B^\alpha{}_\beta{}^\mu)$ where $\bfB := \bfA - 
\bfGamma$ and $\bfGamma = (\Gamma^\alpha{}_\beta)$ is a connection matrix with 
components given by $\Gamma^\alpha{}_\beta = \Gamma^\alpha{}_\beta{}^\mu p_\mu 
= g^{\mu\nu} \Gamma^\alpha{}_{\beta\nu} p_\mu$ where 
$\Gamma^\alpha{}_{\beta\mu}$ are the Christoffel symbols with respect to the 
metric \eqref{eq:metric}.  In this picture $\bfA^\mu$ will therefore represent 
a connection type object.  Using the fact that \eqref{eq:lax} is homogeneous 
and quadratic in the momenta we can write the Lax pair equation on the 
configuration space as
\begin{equation}\label{eq:laxconf}
   \partial_{(\mu}\bfL_{\nu)} = \Gamma^\lambda{}_{\mu\nu}\bfL_\lambda
                               + \bigl[\bfL_{(\mu}\, , \bfA_{\nu)}\bigr] \ .
\end{equation}
The covariant derivative of a third rank tensor $\bfT_\mu = 
(T^\alpha{}_{\beta\mu})$ can be written in matrix form as $\nabla_\mu\bfT_\nu 
= \partial_\mu\bfT_\nu + \rbigl[\bfGamma_\mu, \bfT_\nu\rbigr] - 
\Gamma^\lambda{}_{\mu\nu} \bfT_\lambda$.  Using this relation to replace the 
partial derivative in equation \eqref{eq:laxconf} by the covariant derivative 
we obtain the covariant Lax pair equation \cite{rosquist:lax,rosquist:rmatrix}
\begin{equation}\label{eq:lax_cov}
   \nabla_{(\mu}\bfL_{\nu)} = [\bfL_{(\mu}, \bfB_{\nu)}] \ ,
\end{equation}
where $\bfL_\mu = g_{\mu\nu}\bfL^\nu$, $\bfB_\mu = g_{\mu\nu}\bfB^\nu$ and 
$\nabla_\mu$ represents the covariant derivative associated to the connection 
given by $\Gamma^\alpha{}_{\beta\mu}$.  It is worth pointing out that the 
entire structure of the Lax pair equation \eqref{eq:lax} follows from the 
symmetrized covariant derivative in the left hand side of \eqref{eq:lax_cov}.  
This is true even if $\bfB=0$ so that the right hand side of 
\eqref{eq:lax_cov} vanishes since one can still have a nontrivial Lax pair 
$\bfL$ and $\bfA = \bfGamma$.

\section{The $r$-tensor and the covariant $r$-matrix equation}

Given that the Lax matrices are elements of a vector space $\Vscr$ the 
$r$-matrix is defined on the Kronecker product space $\Vscr \otimes \Vscr$.  
If $\bfX = (X^\alpha{}_\beta)$ and $\bfY = (Y^\alpha{}_\beta)$ are elements of 
$\Vscr$ then their Kronecker product has components $(\bfX \otimes 
\bfY)^{\alpha\mu}{}_{\beta\nu} = X^\alpha{}_\beta Y^\mu{}_\nu$.  Any element 
$\bfX$ in $\Vscr$ can be regarded as an element in $\Vscr \otimes \Vscr$ by 
taking its left or right Kronecker product with the unit matrix.  For this we 
use a notation with space indicators placed within parentheses on top of the 
symbols
\begin{equation}\label{eq:notation2}
   \bs{X}{1} := \bfX\otimes\id \ ,\qquad \bs{X}{2} := \id\otimes\bfX \ .
\end{equation}
Also, if $\bfu$ is an element of $\Vscr\otimes\Vscr$ with components $u^{ij}$ 
with respect to a basis $\bfE_i$ of $\Vscr$ ($i,j,\ldots = 1,2,\ldots,d^2$) we 
write
\begin{equation}\label{eq:notation3}
   \bs{u}{12} = u^{ij}\bfE_i\otimes\bfE_j \ ,\qquad
   \bs{u}{21} = u^{ji}\bfE_i\otimes\bfE_j \ .
\end{equation}
The $r$-matrix equation can then be written \cite{bv:hamlax}
\begin{equation}\label{eq:req}
   \{\bs{L}{1},\bs{L}{2}\} = [\bs{r}{12},\bs{L}{1}] -
                             [\bs{r}{21},\bs{L}{2}] \ .
\end{equation}
We wish to write this equation in covariant form in analogy with the covariant 
formulation of the Lax pair equation.  Computing the Poisson bracket we find 
that the left hand side becomes
\begin{equation}\label{eq:lcomm}
     \{\bs{L}1, \bs{L}2\} =
             \rbigl(-\bsL1{\mu} \del_\mu\bsL2{\nu}
                + \bsL2{\mu} \del_\mu\bsL1{\nu} \rbigr) p_\nu
                          = p_\nu \bfL^\mu \!\wedge \del_\mu\bfL^\nu \ ,
\end{equation}
where we have introduced an exterior Kronecker product, $\bfX\wedge\bfY = 
\bfX\otimes\bfY - \bfY\otimes\bfX$.  Since the left hand side of 
\eqref{eq:req} is linear in the momenta the same must be true for the right 
hand side.  The Lax matrix itself being linear in the momenta it is therefore 
natural to demand that the $r$-matrix is momentum independent.  We can then 
write the $r$-matrix equation on the configuration space as
\begin{equation}\label{eq:rm_config}
   \bfL^\mu \wedge \del_\mu\bfL^\nu
    = [\bs{L}{1}{}^\nu, \bs{r}{12}] - [\bs{L}{2}{}^\nu, \bs{r}{21}] \ .
\end{equation}
Using the formula $\nabla_\mu\bfT^\nu = \partial_\mu\bfT^\nu + 
\rbigl[\bfGamma_\mu, \bfT^\nu\rbigr] + \Gamma^\nu{}_{\lambda\mu} \bfT^\lambda$ 
to rewrite \eqref{eq:rm_config} in terms of covariant derivatives we obtain
\begin{equation}\label{eq:rm_cov}
 \bfL^\mu \wedge \nabla_\mu\bfL^\nu
  = [\bs{L}{1}{}^\nu, \bs{\Rscr}{12}] - [\bs{L}{2}{}^\nu, \bs{\Rscr}{21}] \ ,
\end{equation}
where $\bfRscr := \bfr + \bfs$ and $\bfs := \bfGamma_\lambda \otimes 
\bfL^\lambda$.  This equation is covariant if we interpret $\bfRscr$ as a 
tensor, the $r$-\emph{tensor}.  We will therefore refer to \eqref{eq:rm_cov} 
as the covariant $r$-matrix equation.  For a given solution of 
\eqref{eq:rm_cov} the $r$-matrix itself is given by the relation
\begin{equation}\label{eq:rm}
   \bfr = \bfRscr - \bfGamma_\lambda\otimes\bfL^\lambda \ ,\qquad 
   r^{\alpha\mu}{}_{\beta\nu} = \Rscr^{\alpha\mu}{}_{\beta\nu} - 
   \Gamma^\alpha{}_{\beta\lambda} L^\mu{}_\nu{}^\lambda \ .
\end{equation}
The covariant formulation given here sheds new light on the $r$-matrix 
equation.  In particular, the entire structure of the $r$-matrix equation can 
be derived from the left hand side of \eqref{eq:rm_cov} ({\em cf.} the 
discussion above of the Lax pair equation).  This is because $\bfRscr$ being a 
tensor we may in principle have solutions of \eqref{eq:rm_cov} with 
$\bfRscr=0$ so that the right hand side is zero.  We can then still have a 
nontrivial $r$-matrix given by $\bfr = -\bfGamma_\lambda\otimes\bfL^\lambda$.  
It is also of interest that the $r$-matrix in this formulation is independent 
of the momenta.  It should be noted however that momentum dependence can be 
introduced by a canonical transformation which destroys the geometric form of 
the Hamiltonian ({\em cf.} \cite{rg:intspacetimes}).

\section{Transforming the $r$-matrix}

Another and important consequence of the geometric structure of the $r$-matrix 
given here is that the relation \eqref{eq:rm} can be used to derive the 
transformation properties of the $r$-matrix under coordinate transformations.  
Loosely speaking it follows from \eqref{eq:rm} that the $r$-matrix is somewhat 
of a hybrid object; it is neither a tensor nor a connection in this 
formulation.  More precisely, it carries one connection type index and three 
tensorial indices.

To write down the transformation formula for the $r$-matrix we use relation 
\eqref{eq:rm} together with the known transformation properties of $\bfRscr$,
$\bfGamma_\mu$ and $\bfL^\mu$.  A general coordinate transformation is defined 
by a matrix $\bfU = (U^{\mu'}{}_\mu)$ and its inverse $\bfU^{-1} = 
(U^\mu{}_{\mu'})$ with components given by
\begin{equation}
   U^{\mu'}{}_\mu = \frac{\del q^{\mu'}}{\del q^\mu} \qquad\qquad
   U^\mu{}_{\mu'} = \frac{\del q^\mu}{\del q^{\mu'}} \ .
\end{equation}
Then since $\bfRscr$ and $\bfL^\mu$ are tensors they transform according to 
the formulas
\begin{equation}\label{eq:tenstrans3}
   {\bfL'}^{\mu'} = U^{\mu'}{}_\mu \bfU\bfL^\mu\bfU^{-1} \ ,\qquad
   \bfRscr' = \bfw \bfRscr \bfw^{-1} \ ,
\end{equation}
where $\bfw = \bfU \otimes \bfU$ while $\bfGamma_\mu$ being a connection 
transforms as
\begin{equation}\label{eq:gmatder1}
   \bfGamma'_{\mu'} = U^\mu{}_{\mu'}\bfU\bfGamma_\mu \bfU^{-1}
        + \bfU \,\del_{\mu'}\bfU^{-1} \ .
\end{equation}
Using \eqref{eq:rm} we then obtain the transformation for the 
$r$-matrix as
\begin{equation}\label{eq:rmtrans}
   \bfr' = \bfw\bfr\bfw^{-1}
            + (\del_\mu\bfU\otimes \bfU\bfL^\mu)\bfw^{-1}
         = \bfw(\bfr + \bfU^{-1}\del_\mu\bfU\otimes\bfL^\mu)\bfw^{-1} \ ,
\end{equation}
Although this formula has been derived under the assumptions of the geometric 
framework used in this paper, it is nevertheless valid regardless of those 
assumptions.  In fact, it is straight forward to derive it directly from the 
general form \eqref{eq:req} of the $r$-matrix equation.  Essentially the same 
equation was recently found in that way in \cite{hy:rmatrix}.  However, the
formulation given here has the advantage of yielding a geometric explanation 
of the transformation formula.

\section{The generalized dynamical classical Yang-Baxter equation}

In this paragraph we discuss the \cybe\ in the generalized dynamical form 
which arises in the geometric framework.  Since the left hand side of the 
$r$-matrix equation \eqref{eq:req} involves a Poisson bracket, the right hand 
side is subject to certain restrictions.  It is obvious that the antisymmetry 
property of the Poisson bracket is automatically incorporated in the structure 
of the right hand side.  However, it turns out that the Jacobi identity leads 
to nontrivial restrictions.  To write down the Jacobi identity in the matrix 
formalism we need to extend the notation \eqref{eq:notation2} to the Kronecker 
product space $\Vscr\otimes\Vscr \otimes\Vscr$
\begin{equation}
   \bs{X}{1} := \bfX\otimes\id\otimes\id \ ,\qquad
   \bs{X}{2} := \id\otimes\bfX\otimes\id \ ,\qquad
   \bs{X}{3} := \id\otimes\id\otimes\bfX \ .
\end{equation}
Although the notation then becomes ambiguous (for example $\bs{X}{1}$ can be 
an element in both $\Vscr\otimes\Vscr$ and $\Vscr\otimes\Vscr \otimes\Vscr$) 
it is commonly used in the literature and does not lead to confusion as long 
as we keep in mind in which space we are working. We also extend the notation
\eqref{eq:notation3} in a similar way by forming appropriate Kronecker 
products with the identity matrix in $\Vscr$, {\em e.g.} $\bs{u}{13} = 
u^{ij} \bfE_i\otimes\id \otimes\bfE_j$.

A straight forward but somewhat tedious calculation using \eqref{eq:req} then 
gives \cite{bv:hamlax}
\begin{equation}\label{eq:jacobi}
    \{\{\bs{L}1, \bs{L}2\}, \bs{L}{3}\} + \{\{\bs{L}2, \bs{L}3\}, \bs{L}{1}\}
  + \{\{\bs{L}3, \bs{L}1\}, \bs{L}{2}\}
 = [\bs{f}{123}, \bs{L}1] + [\bs{f}{231}, \bs{L}2]
                          + [\bs{f}{312}, \bs{L}{3}] \ ,
\end{equation}
where $\bs{f}{abc} := \bs{r}{abc} + \bs{q}{abc}$ and we are using the 
notation
\begin{subequations}
 \begin{align}
   \bs{r}{abc} &:= [\bs{r}{ab}, \bs{r}{ac}] + [\bs{r}{ab}, \bs{r}{bc}]
                 +[\bs{r}{cb}, \bs{r}{ac}] \ ,\label{eq:rabc}\\
   \bs{q}{abc} &:= \{\bs{L}{b}, \bs{r}{ac}\} - \{\bs{L}{c}, \bs{r}{ab}\} \ ,
                   \label{eq:q1}
 \end{align}
\end{subequations}
where the triplet $(a,b,c)$ is a permutation of $(1,2,3)$. From 
\eqref{eq:jacobi} it follows that the Jacobi identity implies the relation
\begin{equation}\label{eq:dybe}
   [\bs{f}{123}, \bs{L}1] + [\bs{f}{231}, \bs{L}2]
                          + [\bs{f}{312}, \bs{L}{3}] = 0 \ .
\end{equation}
This relation may be interpreted as a constraint on the $r$-matrix imposed by 
the Jacobi identity.  Trying to write \eqref{eq:dybe} in covariant form we 
first note that if the $r$-matrix has no momentum dependence then 
\eqref{eq:q1} becomes
\begin{equation}
     \bs{q}{abc} =   \bs{L}{c}{}^\mu\,\del_\mu\!\bs{r}{ab}
                    - \bs{L}{b}{}^\mu\,\del_\mu\!\bs{r}{ac} \ .
\end{equation}
The next step is to use \eqref{eq:rm} to replace the $r$-matrix by the 
$r$-tensor, $\bfRscr$.  We must also replace the partial derivative by the 
covariant derivative.  For a fourth rank tensor $\bfu = 
(u^{\alpha\mu}{}_{\beta\nu})$ (such as the $r$-tensor) the covariant 
derivative can be written in matrix form as
\begin{equation}\label{eq:sder2}
     \nabla_\lambda\bs{u}{ab} = \del_\lambda\bs{u}{ab}
       + [{\bs{\Gamma}a}_\lambda + {\bs{\Gamma}b}_\lambda, \bs{u}{ab}] \ .
\end{equation}
After another rather long calculation the result is \cite{rosquist:rmatrix}
\begin{equation}\label{eq:fabc}
   \bs{f}{abc} = \bs{\Rscr}{abc} + \bs{L}{c}{}^\mu\nabla_\mu\bs{\Rscr}{ab}
                                 - \bs{L}{b}{}^\mu\nabla_\mu\bs{\Rscr}{ac}
                + \bs{L}{b}{}^\mu \bs{L}{c}{}^\nu \bs{F}{a}_{\mu\nu} \ ,
\end{equation}
where the $\bfF_{\mu\nu}$ are curvature matrices representing the components 
of the Riemann tensor according to
\begin{equation}
   \bfF_{\mu\nu} = (R^\alpha{}_{\beta\mu\nu})
                 = 2\partial_{[\mu}\bfGamma_{\nu]}
                  + [\bfGamma_\mu, \bfGamma_\nu] \ .
\end{equation}

The relation \eqref{eq:fabc} together with \eqref{eq:dybe} represents the 
covariant form of the restriction on the $r$-matrix implied by the Jacobi 
identity.  It can therefore be viewed as a covariant generalized dynamical 
\cybe.  One often imposes additional restrictions which serve as sufficient 
conditions to satisfy the Jacobi identity constraint \eqref{eq:dybe}.  The 
most general sufficient conditions so far were given in 
\cite{sklyanin:dynamicalr}.  A commonly used restriction is to assume that the 
$r$-matrix is purely numerical and satisfies the unitarity condition 
$\bs{r}{12} = -\bs{r}{21}$.  Then any solution of the original \cybe\ (note 
the difference in the space indicators in the last term compared to 
\eqref{eq:rabc})
\begin{equation}
   [\bs{r}{ab}, \bs{r}{ac}] + [\bs{r}{ab}, \bs{r}{bc}]
                            + [\bs{r}{ac}, \bs{r}{bc}] = 0 \ ,
\end{equation}
satisfies the Jacobi identity constraint \eqref{eq:dybe}.  In the geometric 
context such restrictions can be regarded as gauge conditions leading to 
noncovariant Yang-Baxter equations.  Judging from the form of \eqref{eq:fabc} 
there seems to be no simple covariant restriction on $\bfRscr$ which solves 
\eqref{eq:dybe} for general metrics $g_{\mu\nu}$.  If the geometry is flat so 
that $\bfF_{\mu\nu} =0$, then the covariant condition $\bfRscr=0$ solves 
\eqref{eq:dybe}, leading to the expression $\bfr = -\bfGamma_\lambda \otimes 
\bfL^\lambda$ for the $r$-matrix.  It should be noted that the flat case is 
far from trivial.  For example, suppose $\bfL=\bfL^\mu p_\mu$ is a Lax matrix 
for the flat geometry and $\bfRscr$ is an accompanying $r$-tensor (possibly 
zero).  Define a new geometry by $h^{\mu\nu} := \Tr(\bfL^\mu \bfL^\nu)$.  In 
general, $h^{\mu\nu}$ is a non-flat metric and the corresponding Hamiltonian 
system is therefore integrable but physically different from the original flat 
system.

\section{Outlook}
The formalism introduced in this paper by itself provides a new view of the 
nature of the classical $r$-matrix.  However, to obtain additional insights 
into the properties of the classical $r$-matrix it would be useful to have 
explicit examples of $r$-tensors.  In fact it seems natural to try to find 
such an example by looking for an $r$-tensor for the geometric version of the 
non-periodic Toda lattice given in \cite{rg:intspacetimes}.  Given that such 
an $r$-tensor exists, the corresponding $r$-matrix would depend on the 
configuration variables but not on the momenta.  However, going back to 
the original non-geometric formulation using the inverse of the canonical 
transformation given in \cite{rg:intspacetimes}, the $r$-matrix would seem to 
pick up a momentum dependence.  Presumably that momentum dependence could then 
be removed by using the freedom in the construction of the $r$-matrix recently 
investigated by Braden \cite{braden:rmgeninv}.


\end{document}